# ANALYSIS OF REINFORCED CONCRETE SHELLS
# WITH TRANSVERSE SHEAR FORCES

By Mauro Schulz[1] and Maria Paola Santisi d'Avila[2]

**ABSTRACT:**

This research investigates the simultaneous effect of in-plane and transverse loads in reinforced concrete shells. The infinitesimal shell element is divided into layers (with triaxial behavior) that are analyzed according to the smeared rotating crack approach. The set of internals forces includes the derivatives of the in-plane components. The corresponding generalized strains are determined using an extension of the equivalent section method, valid for shells. The formulation yields through-the-thickness distributions of stresses and strains and the spatial orientation of the concrete struts. Although some simplifications are necessary to establish a practical first-order approximation, higher-order solutions could be developed. Despite the fact that constitutive matrices are not symmetric, because of the tension-softening formulation, the equilibrium and compatibility conditions are satisfied, the stiffness derivatives are explicitly calculated and the algorithms show good convergence. The formulation predicts results that agree with experimental data obtained by other researchers. Although comparative analysis with additional experimental data is still necessary, the proposed theory provides a promising solution for the design of reinforced concrete shells.

**Keywords:** Shell structures; reinforced concrete; triaxial stress; shear strength; structural design.

[1]Professor, Dept. of Civil Engineering, Universidade Federal Fluminense, Niterói, RJ, Brazil, 24220-900
[2]Ph.D., Laboratoire Central des Ponts et Chaussées, Paris, France, 75732

**INTRODUCTION**

Reinforced concrete shell structures are widely employed because they provide both outstanding performance and architectural beauty. The design of reinforced concrete shells involves the determination of reinforcement that is not necessarily arranged in the direction of the principal internal forces. The mechanical model must consider the entire set of in-plane and out-of-plane forces, bending and twisting moments, as well as the spatial distribution of stresses and strains.

After early investigations regarding the analysis of reinforced concrete plates and shells, Falconer (1956) establishes the equilibrium equations of a plane stress element, considering the axial forces of the reinforcement, placed in two different orientations, and a compression field of concrete stresses, the direction of which remains statically indeterminate.

Using the plasticity approach, Nielsen (1964) defines design equations for membrane elements. Wood (1968) and Armer (1968) organize a similar procedure for the flexural design of reinforced concrete slabs. Baumann (1972a; b) considers membrane and flexural forces simultaneously, by distributing the internal forces between two plate elements at the upper and lower faces of the shell. Each plate is analyzed separately, using the compatibility approach. Baumann evaluates the direction of the struts of each membrane element by the principle of minimum work, arriving at a linear-elastic approximation of the general equation later proposed by Mitchell and Collins (1974). The total reinforcement is minimized taken into account the requirement of minimum secondary reinforcement. The shell design procedure recommended by CEB and FIP (1982) applies the two-plate approximation together with the plasticity approach, to arrive at a simple design method that is equivalent to the compatibility formulation, where the contribution of the secondary reinforcement is neglected.

The shell formulations discussed above are extensively applied, but they do have the following limitations: the thickness and the lever arm of the upper and lower membrane elements are not explicitly calculated, the materials are not represented by nonlinear stress-strain relationships and the beneficial effects of the compression reinforcement are not taken into account. Schnobrich (1977) reports early applications of multi-layered models in finite element analysis of reinforced concrete structures. Multi-layered models are used by Schulz (1984) and Kirschner and Collins (1986) to verify the punctual response of reinforced concrete shells to in-plane forces, and twisting and flexural moments. The assumption that plane sections remain plane in any direction yields the biaxial strains of each layer. The corresponding stresses are obtained by applying nonlinear constitutive models.

The design of shells for transverse shear forces involves additional considerations. Schulz (1988) and Marti (1990) assume that shear stresses are distributed according to the corresponding principal direction, since no shear force is acting perpendicularly. The principal shear force, which is determined by the square root of the sum of the squares of the components in any system of coordinates, is used as a reference for simplified design formulas.

Analytical models which consider flexural and shear effects simultaneously are also investigated in the literature. Kirschner and Collins (1986) include out-of-plane shear by dividing the reinforced concrete shell into three-dimensional elements. Three-dimensional constitutive relationships are applied assuming a perfectly uniform stress distribution of transverse shear stresses in the core of the shell. Adebar (1989) and Adebar and Collins (1994) analyze the punctual response of a shell element as subjected to in-plane loads only. The simplified model is corrected by complementary in-plane forces, which are evaluated by assuming a shear stress value at the midpoint of the shell and applying three-dimensional constitutive relationships at this point. Polak and Vecchio (1993) investigate transverse shear in reinforced concrete shells using a finite element based on Reissner-Mindlin plate theory. By assuming that plane sections remain plane but not necessarily normal to the mid-surface, transverse shear strains are considered constant along the thickness of the shell.

The last models represent a significant advance, but the evaluation of transverse shear stresses and strains can be improved. The present research investigates a mechanical model for reinforced concrete shells considering simultaneous in-plane and transverse shear forces. The infinitesimal shell element is divided along the thickness into infinitesimal three-dimensional elements, which respect three-dimensional constitutive relationships. The implementation is based on the smeared rotating crack approach and the shear stiffness terms are deduced accordingly. The theory applies to shells, the equivalent section method for beams, proposed by Diaz (1980) and Diaz and Schulz (1981). Through-the-thickness distributions of stresses and strains are established considering equilibrium and compatibility conditions. The formulation yields the spatial orientation of the concrete struts and the transverse shear stresses are not necessarily oriented according to the principal shear force. Some simplifications are necessary to establish a first-order approximation, which is recommended for design applications. However, it is shown how to develop higher-order solutions for a theoretically precise response.

**SIMPLIFYING HYPOTHESES**

The discussion here is limited to small displacements. Disturbed states of deformation at boundaries and load application points are not taken into account. The following hypotheses are introduced at the outset:

1. Eventual cracks are considered uniformly distributed and concrete stresses and strains are stated as continuous and derivable functions. The spatial orientation of the smeared struts varies along the element thickness.

2. No slip is considered between concrete and reinforcing bars. Increments of steel and concrete strains are assumed to be equal on an average basis.

3. The principal directions of concrete stresses and strains are considered coincident.

4. Boundary and volume forces are not taken into account in the interest of simplifying the formulation.

5. The resultant of concrete and steel stresses in the $z$-direction is small and, hence, neglected.

6. In-plane strains are linearly distributed along the thickness (generalized Bernoulli's hypothesis).

**REINFORCED CONCRETE THREE-DIMENSIONAL ELEMENT**

A shell element is presented in Figure 1. The $z$-axis is normal to the mid-plane of the shell and two-way reinforcement layers are placed according to $x$- and $y$-axes. Although this steel distribution is frequent and simplifies the formulation, it is possible to consider skew and multidirectional reinforcement through additional coordinate transformations. The transverse shear reinforcement is considered whenever present. The shell element, with infinitesimal dimensions $dx\,dy$ and finite thickness $t$, is divided into three-dimensional elements which have infinitesimal dimensions $dx\,dy\,dz$ (Figure 2). Tensile strains and stresses are considered positive.

Concrete is assumed to be a continuous and uniform medium (hypothesis 1). The concrete strain vector $\boldsymbol{\varepsilon}$ is defined by

$$\boldsymbol{\varepsilon} = \begin{bmatrix} \varepsilon_x & \varepsilon_y & \gamma_{xy} & \gamma_{xz} & \gamma_{yz} & \varepsilon_z \end{bmatrix}^T \tag{1}$$

The normal strains in $x$-, $y$- and $z$-directions are $\varepsilon_x$, $\varepsilon_y$ and $\varepsilon_z$. The shear strains are denoted by $\gamma_{xy}$, $\gamma_{xz}$ and $\gamma_{yz}$. Concrete strains can also be represented in tensor form by

$$\mathcal{E} = \begin{bmatrix} \varepsilon_x & \varepsilon_{xy} & \varepsilon_{xz} \\ \varepsilon_{xy} & \varepsilon_y & \varepsilon_{yz} \\ \varepsilon_{xz} & \varepsilon_{yz} & \varepsilon_z \end{bmatrix} \tag{2}$$

where $\gamma_{ij} = 2\varepsilon_{ij}$. According to hypothesis 2, the slip between steel bars and concrete is neglected. The reinforcement strain varies according to the average strain of the surrounding concrete. The steel strains $\varepsilon_{sx}$, $\varepsilon_{sy}$ and $\varepsilon_{sz}$, respectively in $x$-, $y$- and $z$-directions, are determined by

$$\varepsilon_{sx} = \varepsilon_x + \varepsilon_{sx0} \qquad \varepsilon_{sy} = \varepsilon_y + \varepsilon_{sy0} \qquad \varepsilon_{sz} = \varepsilon_z + \varepsilon_{sz0} \tag{3a-c}$$

The terms $\varepsilon_{sx0}$, $\varepsilon_{sy0}$ and $\varepsilon_{sz0}$ represent residual strains of pre-tensioned or bonded post-tensioned tendons, which are calculated considering tensioning operations, mobilized loading and prestressing losses.

The concrete stress vector, in the $xyz$ coordinate system, is defined by

$$\boldsymbol{\sigma} = \begin{bmatrix} \sigma_x & \sigma_y & \tau_{xy} & \tau_{xz} & \tau_{yz} & \sigma_z \end{bmatrix}^T \tag{4}$$

Concrete stresses can be represented in tensor form by

$$S = \begin{bmatrix} \sigma_x & \tau_{xy} & \tau_{xz} \\ \tau_{xy} & \sigma_y & \tau_{yz} \\ \tau_{xz} & \tau_{yz} & \sigma_z \end{bmatrix} \tag{5}$$

The steel stress vector $\boldsymbol{\sigma}_s$ is expressed by

$$\boldsymbol{\sigma}_s = \begin{bmatrix} \sigma_{sx} & \sigma_{sy} & 0 & 0 & 0 & \sigma_{sz} \end{bmatrix}^T \tag{6}$$

where $\sigma_{sx}$ and $\sigma_{sy}$ are stresses of the longitudinal reinforcement and $\sigma_{sz}$ represents the transverse reinforcement stress.

The analogous stress vector $\mathbf{s}$, which combines the contributions of both concrete and reinforcement, is defined by

$$\mathbf{s} = \begin{bmatrix} s_x & s_y & s_{xy} & s_{xz} & s_{yz} & s_z \end{bmatrix}^T = \boldsymbol{\sigma} + \boldsymbol{\rho}_s \boldsymbol{\sigma}_s \tag{7}$$

where matrix $\boldsymbol{\rho}_s$ of the steel ratios is

$$\boldsymbol{\rho}_s = \begin{bmatrix} \rho_{sx} & & & & & \mathbf{0} \\ & \rho_{sy} & & 0 & & \\ & & 0 & & & \\ & & & 0 & & \\ & \mathbf{0} & & & 0 & \\ & & & & & \rho_{sz} \end{bmatrix} \tag{8}$$

The non-dimensional terms $\rho_{sx}$, $\rho_{sy}$ and $\rho_{sz}$ represent the ratios of steel area, respectively in $x$-, $y$- and $z$-directions, over the corresponding concrete areas $dy\,dz$, $dz\,dx$ and $dx\,dy$.

Although several approaches for modeling the nonlinear behavior of reinforced concrete could be adopted, a simple hyperelastic stress-strain relationship is considered adequate for the analysis of reinforced concrete elements under monotonic loadings. According to hypothesis 3, shear stresses and strains are equal to zero in the principal coordinate system. The concrete stress-strain relationship in the principal coordinates is expressed by

$$\sigma_1 = \sigma_1(\varepsilon_1, \varepsilon_2, \varepsilon_3) \qquad \sigma_2 = \sigma_2(\varepsilon_1, \varepsilon_2, \varepsilon_3) \qquad \sigma_3 = \sigma_3(\varepsilon_1, \varepsilon_2, \varepsilon_3) \qquad (9)a\text{-}c$$

The definition of the constitutive law in the principal coordinate system reduces the number of variables and simplifies the constitutive formulation, but demands coordinate transformations. The transformation rule of second order tensors yields

$$\mathcal{E}_{123} = \mathbf{\Phi}^T \mathcal{E} \mathbf{\Phi} \qquad (10)$$

where $\mathcal{E}_{123}$ is the strain tensor in the principal coordinate system $(x_1 x_2 x_3)$. The corresponding rotation matrix $\mathbf{\Phi}$ is expressed by

$$\mathbf{\Phi} = \begin{bmatrix} \boldsymbol{\phi}_1 & \boldsymbol{\phi}_2 & \boldsymbol{\phi}_3 \end{bmatrix} = \begin{bmatrix} \phi_{1x} & \phi_{2x} & \phi_{3x} \\ \phi_{1y} & \phi_{2y} & \phi_{3y} \\ \phi_{1z} & \phi_{2z} & \phi_{3z} \end{bmatrix} \qquad (11)$$

The terms $\phi_{ix}$, $\phi_{iy}$ and $\phi_{iz}$ are the direction cosines of the $x_i$-coordinate axis with respect to $x$-, $y$-and $z$-directions (Figure 3). The same transformation can be expressed in vector form by

$$\boldsymbol{\varepsilon}_{123} = \mathbf{T} \boldsymbol{\varepsilon} \qquad (12)$$

where $\boldsymbol{\varepsilon}_{123}$ is the principal strain vector. According to Cook, Malkus and Plesha (1989), the transformation matrix $\mathbf{T}$ is defined by

$$\mathbf{T} = \begin{bmatrix} \phi_{1x}^2 & \phi_{1y}^2 & \phi_{1x}\phi_{1y} & \phi_{1x}\phi_{1z} & \phi_{1y}\phi_{1z} & \phi_{1z}^2 \\ \phi_{2x}^2 & \phi_{2y}^2 & \phi_{2x}\phi_{2y} & \phi_{2x}\phi_{2z} & \phi_{2y}\phi_{2z} & \phi_{2z}^2 \\ 2\phi_{1x}\phi_{2x} & 2\phi_{1y}\phi_{2y} & \phi_{1x}\phi_{2y}+\phi_{1y}\phi_{2x} & \phi_{1x}\phi_{2z}+\phi_{2x}\phi_{1z} & \phi_{1y}\phi_{2z}+\phi_{2y}\phi_{1z} & 2\phi_{1z}\phi_{2z} \\ 2\phi_{1x}\phi_{3x} & 2\phi_{1y}\phi_{3y} & \phi_{1x}\phi_{3y}+\phi_{3x}\phi_{1y} & \phi_{1z}\phi_{3x}+\phi_{3z}\phi_{1x} & \phi_{1y}\phi_{3z}+\phi_{3y}\phi_{1z} & 2\phi_{1z}\phi_{3z} \\ 2\phi_{2x}\phi_{3x} & 2\phi_{2y}\phi_{3y} & \phi_{2x}\phi_{3y}+\phi_{3x}\phi_{2y} & \phi_{2z}\phi_{3x}+\phi_{3z}\phi_{2x} & \phi_{2y}\phi_{3z}+\phi_{3y}\phi_{2z} & 2\phi_{2z}\phi_{3z} \\ \phi_{3x}^2 & \phi_{3y}^2 & \phi_{3x}\phi_{3y} & \phi_{3x}\phi_{3z} & \phi_{3y}\phi_{3z} & \phi_{3z}^2 \end{bmatrix} \quad (13)$$

Equation (10) defines a spectral decomposition. The principal strain tensor $\mathcal{E}_{123}$, expressed by

$$\mathcal{E}_{123} = \begin{bmatrix} \varepsilon_1 & 0 & 0 \\ 0 & \varepsilon_2 & 0 \\ 0 & 0 & \varepsilon_3 \end{bmatrix} \quad (14)$$

is a spectral matrix having as diagonal elements the eigenvalues of $\mathcal{E}$. The columns of the modal matrix $\mathbf{\Phi}$ are the eigenvectors of $\mathcal{E}$. The solution of the eigenproblem (10) yields the principal strains $\varepsilon_1$, $\varepsilon_2$ and $\varepsilon_3$ and the rotation matrix $\mathbf{\Phi}$. The principal stresses $\sigma_1$, $\sigma_2$ and $\sigma_3$ are determined by the constitutive relationships (9). The principal stress tensor $S_{123}$ corresponds to

$$S_{123} = \begin{bmatrix} \sigma_1 & 0 & 0 \\ 0 & \sigma_2 & 0 \\ 0 & 0 & \sigma_3 \end{bmatrix} \quad (15)$$

According to hypothesis 3, the concrete stress components in $xyz$ coordinates are determined by either of the following transformations:

$$S = \mathbf{\Phi} S_{123} \mathbf{\Phi}^T \qquad \mathbf{\sigma} = \mathbf{T}^T \mathbf{\sigma}_{123} \quad (16)a,b$$

The procedure defined by the spectral decomposition (10), the constitutive equations (9) and the coordinate transformation (16) is valid for cracked and uncracked concrete. The solution of nonlinear problems demands incremental equations. The increment $\Delta S_{123}$ of the stress tensor is

expressed by

$$\Delta S_{123} = \begin{bmatrix} \Delta\sigma_1 & \Delta\tau_{12} & \Delta\tau_{13} \\ \Delta\tau_{12} & \Delta\sigma_2 & \Delta\tau_{23} \\ \Delta\tau_{13} & \Delta\tau_{23} & \Delta\sigma_3 \end{bmatrix} \qquad (17)$$

Matrix $\Delta S_{123}$ is not necessarily diagonal. Considering that the constitutive functions (9) are sufficient to relate any state of strain to the corresponding stresses, they must also be sufficient to define the tangent constitutive matrix. The incremental equation is expressed in the principal coordinate system $(x_1 x_2 x_3)$ by

$$\begin{aligned} S_{123} + \Delta S_{123} &= (\mathbf{I} + \Delta\mathbf{\Phi})(S_{123} + \Delta\overline{S}_{123})(\mathbf{I} + \Delta\mathbf{\Phi})^T = \\ &= S_{123} + \Delta\mathbf{\Phi}\, S_{123}\, \mathbf{I}^T + \mathbf{I}\,\Delta\overline{S}_{123}\, \mathbf{I}^T + \mathbf{I}\, S_{123}\, \Delta\mathbf{\Phi}^T \end{aligned} \qquad (18)$$

where $S_{123} + \Delta S_{123}$ are updated total stresses, $S_{123} + \Delta\overline{S}_{123}$ are updated principal stresses and $\mathbf{I} + \Delta\mathbf{\Phi}$ defines the coordinate transformation from the former to the updated principal directions. Matrix $\mathbf{I} + \Delta\mathbf{\Phi}$ is a rotation matrix for an infinitesimal increment. Matrices $\Delta\overline{S}_{123}$ and $\Delta\mathbf{\Phi}$ are defined by

$$\Delta\overline{S}_{123} = \begin{bmatrix} \Delta\sigma_1 & 0 & 0 \\ 0 & \Delta\sigma_2 & 0 \\ 0 & 0 & \Delta\sigma_3 \end{bmatrix} \qquad \Delta\mathbf{\Phi} = \begin{bmatrix} 0 & -\Delta\phi_3 & \Delta\phi_2 \\ \Delta\phi_3 & 0 & -\Delta\phi_1 \\ -\Delta\phi_2 & \Delta\phi_1 & 0 \end{bmatrix} \qquad (19)a,b$$

Using (18) and (19), the stress tensor increment is obtained by

$$\Delta S_{123} = \begin{bmatrix} \Delta\sigma_1 & (\sigma_1-\sigma_2)\Delta\phi_3 & (\sigma_3-\sigma_1)\Delta\phi_2 \\ (\sigma_1-\sigma_2)\Delta\phi_3 & \Delta\sigma_2 & (\sigma_2-\sigma_3)\Delta\phi_1 \\ (\sigma_3-\sigma_1)\Delta\phi_2 & (\sigma_2-\sigma_3)\Delta\phi_1 & \Delta\sigma_3 \end{bmatrix} \qquad (20)$$

Applying the same procedure, the strain tensor increment is expressed by

$$\Delta \boldsymbol{\mathcal{E}}_{123} = \begin{bmatrix} \Delta\varepsilon_1 & (\varepsilon_1-\varepsilon_2)\Delta\phi_3 & (\varepsilon_3-\varepsilon_1)\Delta\phi_2 \\ (\varepsilon_1-\varepsilon_2)\Delta\phi_3 & \Delta\varepsilon_2 & (\varepsilon_2-\varepsilon_3)\Delta\phi_1 \\ (\varepsilon_3-\varepsilon_1)\Delta\phi_2 & (\varepsilon_2-\varepsilon_3)\Delta\phi_1 & \Delta\varepsilon_3 \end{bmatrix} \quad (21)$$

For arbitrary infinitesimal rotations $\Delta\phi_1$, $\Delta\phi_2$ and $\Delta\phi_3$, equations (20) and (21) yield

$$\Delta\tau_{ij} = \Delta\varepsilon_{ij}\left(\sigma_i - \sigma_j\right)/\left(\varepsilon_i - \varepsilon_j\right) \quad (22)$$

with $i,j = 1,2,3$. The relationship between increments of concrete stresses and strains, respectively denoted by $\Delta\boldsymbol{\sigma}_{123}$ and $\Delta\boldsymbol{\varepsilon}_{123}$, is defined by

$$\Delta\boldsymbol{\sigma}_{123} = \mathbf{E}_{123}\,\Delta\boldsymbol{\varepsilon}_{123} \quad (23)$$

Using (9) and (22), the tangent constitutive matrix $\mathbf{E}_{123}$ is expressed by

$$\mathbf{E}_{123} = \begin{bmatrix} E_{11} & E_{12} & 0 & 0 & 0 & E_{13} \\ E_{21} & E_{22} & 0 & 0 & 0 & E_{23} \\ 0 & 0 & G_{12} & 0 & 0 & 0 \\ 0 & 0 & 0 & G_{31} & 0 & 0 \\ E_{31} & 0 & 0 & 0 & G_{23} & 0 \\ E_{32} & 0 & 0 & 0 & 0 & E_{33} \end{bmatrix} \quad (24)$$

where

$$G_{ij} = \frac{\sigma_j - \sigma_i}{2(\varepsilon_j - \varepsilon_i)} \quad (25)$$

Expression (25), based on the coaxiality of principal stresses and strains, is adopted by Willam, Pramono and Sture (1987), Stevens, Uzumeri and Collins (1987), Schulz (1988) and intensively discussed by Zhu, Hsu and Lee (2001).

Expressions (12), (16) and (23) yield

$$\Delta \boldsymbol{\sigma} = \mathbf{E}\,\Delta \boldsymbol{\varepsilon} \qquad (26)$$

The tangent constitutive matrix $\mathbf{E}$, in the $xyz$ coordinate system, is given by

$$\mathbf{E} = \mathbf{T}^T \mathbf{E}_{123}\, \mathbf{T} \qquad (27)$$

The constitutive equations of the steel reinforcement are expressed by $\sigma_{sx} = \sigma_{sx}(\varepsilon_{sx})$, $\sigma_{sy} = \sigma_{sy}(\varepsilon_{sy})$ and $\sigma_{sz} = \sigma_{sz}(\varepsilon_{sz})$. The relationship between steel stress and strain increments, respectively $\Delta \boldsymbol{\sigma}_s$ and $\Delta \boldsymbol{\varepsilon}$, is defined by

$$\Delta \boldsymbol{\sigma}_s = \mathbf{E}_s\, \Delta \boldsymbol{\varepsilon} \qquad (28)$$

The constitutive matrix $\mathbf{E}_s$ is

$$\mathbf{E}_s = \begin{bmatrix} E_{sx} & & & & \mathbf{0} \\ & E_{sy} & 0 & & \\ & & & 0 & \\ & \mathbf{0} & & 0 & \\ & & & & E_{sz} \end{bmatrix} \qquad (29)$$

Using (7), (26) and (28), the increment of the stress vector $\Delta \mathbf{s}$ is expressed by

$$\Delta \mathbf{s} = \mathbf{C}\, \Delta \boldsymbol{\varepsilon} \qquad (30)$$

where the constitutive matrix $\mathbf{C}$ of the reinforced concrete element is $\mathbf{C} = \mathbf{E} + \boldsymbol{\rho}_s\, \mathbf{E}_s$.

**REINFORCED CONCRETE SHELL ELEMENT**

The in-plane internal forces $N_x$, $N_y$, $N_{xy}$, $M_x$, $M_y$ and $M_{xy}$ (Figure 1) are defined by

$$N_x = \int_{z_A}^{z_B} s_x \, dz \qquad N_y = \int_{z_A}^{z_B} s_y \, dz \qquad N_{xy} = \int_{z_A}^{z_B} s_{xy} \, dz$$

$$M_x = \int_{z_A}^{z_B} s_x z \, dz \qquad M_y = \int_{z_A}^{z_B} s_y z \, dz \qquad M_{xy} = \int_{z_A}^{z_B} s_{xy} z \, dz \qquad (31)a\text{-}f$$

where $z_A$ and $z_B$ correspond to the upper and lower bounds of the shell element (Figure 2). The transverse shear forces $V_x$ and $V_y$ are expressed by

$$V_x = \int_{z_A}^{z_B} s_{xz} \, dz \qquad V_y = \int_{z_A}^{z_B} s_{yz} \, dz \qquad (32)a,b$$

The following differential equilibrium equation, with respect to $z$-axis, is established assuming that no forces are applied on the element (hypothesis 4) and $s_z \cong 0$ (hypothesis 5):

$$s'_{xz} + \dot{s}_{yz} = 0 \qquad (33)$$

The prime, the point and the star respectively denote partial derivatives of a function with respect to $x$, $y$ and $z$ $\left([\ ]' = \partial[\ ]/\partial x, [\ ]\dot{} = \partial[\ ]/\partial y \text{ and } [\ ]^* = \partial[\ ]/\partial z \right)$. Integrating (33) yields

$$V'_x + \dot{V}_y = 0 \qquad (34)$$

The differential equilibrium equations of the reinforced concrete element with respect to $x$- and $y$-axes are expressed by

$$s^*_{xz} = -s'_x - \dot{s}_{xy} \qquad s^*_{yz} = -s'_{xy} - \dot{s}_y \qquad (35)a,b$$

Equations (35) and the boundary conditions

$$s_{xz}(z_A) = s_{yz}(z_A) = s_{xz}(z_B) = s_{yz}(z_B) = 0 \qquad (36)$$

are defined considering hypothesis 4. Integrating (35) yields the shear stresses, as follows:

$$s_{xz}(z) = -\int_{z_A}^{z} \left( s'_x + \dot{s}_{xy} \right) dz \qquad s_{yz}(z) = -\int_{z_A}^{z} \left( s'_{xy} + \dot{s}_y \right) dz \qquad (37)a,b$$

Integrating (35) and itself multiplied by $z$, and substituting (31), (32) and (36) yield

$$\begin{aligned} N'_x + \dot{N}_{xy} &= 0 & M'_x + \dot{M}_{xy} &= V_x \\ N'_{xy} + \dot{N}_y &= 0 & M'_{xy} + \dot{M}_y &= V_y \end{aligned} \qquad (38)a\text{-}d$$

According to (37), the derivatives $s'_x$, $\dot{s}_{xy}$, $s'_{xy}$ and $\dot{s}_y$ are necessary to evaluate the shear stresses. Equation (30) is expanded according to

$$\left[ \begin{array}{c} \Delta \mathbf{s}_n \\ \hline \Delta \mathbf{s}_t \end{array} \right] = \left[ \begin{array}{c|c} \mathbf{C}_{nn} & \mathbf{C}_{nt} \\ \hline \mathbf{C}_{tn} & \mathbf{C}_{tt} \end{array} \right] \left[ \begin{array}{c} \Delta \boldsymbol{\varepsilon}_n \\ \hline \Delta \boldsymbol{\varepsilon}_t \end{array} \right] \qquad (39)$$

where

$$\begin{aligned} \Delta \mathbf{s}_n &= \begin{bmatrix} \Delta s_x & \Delta s_y & \Delta s_{xy} \end{bmatrix}^T & \Delta \mathbf{s}_t &= \begin{bmatrix} \Delta s_{xz} & \Delta s_{yz} & \Delta s_z \end{bmatrix}^T \\ \Delta \boldsymbol{\varepsilon}_n &= \begin{bmatrix} \Delta \varepsilon_x & \Delta \varepsilon_y & \Delta \gamma_{xy} \end{bmatrix}^T & \Delta \boldsymbol{\varepsilon}_t &= \begin{bmatrix} \Delta \gamma_{xz} & \Delta \gamma_{yz} & \Delta \varepsilon_z \end{bmatrix}^T \end{aligned} \qquad (40)a\text{-}d$$

In (39) and (40), $\Delta \mathbf{s}_n$ and $\Delta \boldsymbol{\varepsilon}_n$ are increments of in-plane stresses and strains. Vectors $\Delta \mathbf{s}_t$ and $\Delta \boldsymbol{\varepsilon}_t$ correspond to transverse normal and shear variables.

The derivatives of $s_z$ are neglected according to hypothesis 5 $(s_z \cong 0)$. A first-order approximation is defined by neglecting the first derivatives of shear stresses $s_{xz}$ and $s_{yz}$, which results in

$$\mathbf{s}'_n = \mathbf{D} \boldsymbol{\varepsilon}'_n \qquad \dot{\mathbf{s}}_n = \mathbf{D} \dot{\boldsymbol{\varepsilon}}_n \qquad (41)a,b$$

The $(3 \times 3)$-constitutive matrix $\mathbf{D}$ is determined by $\mathbf{D} = \mathbf{C}_{nn} - \mathbf{C}_{nt} \mathbf{C}_{tt}^{-1} \mathbf{C}_{tn}$. The terms of matrix $\mathbf{D}$ are forces per unit area. Matrix $\mathbf{D}$ is partitioned according to

$$\mathbf{D}^T = \begin{bmatrix} \mathbf{D}_x^T & | & \mathbf{D}_y^T & | & \mathbf{D}_z^T \end{bmatrix}^T \qquad (42)$$

The following equations are deduced from (41) and (42):

$$s'_x = \mathbf{D}_x \boldsymbol{\varepsilon}'_n \qquad s'_{xy} = \mathbf{D}_{xy} \boldsymbol{\varepsilon}'_n \qquad \dot{s}_y = \mathbf{D}_y \dot{\boldsymbol{\varepsilon}}_n \qquad \dot{s}_{xy} = \mathbf{D}_{xy} \dot{\boldsymbol{\varepsilon}}_n \qquad (43)a\text{-}d$$

Equations (37) and (43) yield

$$s_{xz}(z) = -\int_{z_A}^{z} \left( \mathbf{D}_x \boldsymbol{\varepsilon}'_n + \mathbf{D}_{xy} \dot{\boldsymbol{\varepsilon}}_n \right) \mathrm{d}z \qquad s_{xz}(z) = -\int_{z_A}^{z} \left( \mathbf{D}_{xy} \boldsymbol{\varepsilon}'_n + \mathbf{D}_y \dot{\boldsymbol{\varepsilon}}_n \right) \mathrm{d}z \qquad (44)a,b$$

According to the generalized Bernoulli's hypothesis (hypothesis 6), the longitudinal strains $\varepsilon_x$, $\varepsilon_y$ and $\gamma_{xy}$ are linearly interpolated according to

$$\varepsilon_x = e_x + k_x z \qquad \varepsilon_y = e_y + k_y z \qquad \gamma_{xy} = e_{xy} + k_{xy} z \qquad (45)\ a\text{-}c$$

where $e_x$, $e_y$ and $e_{xy}$ are strains at $z=0$ and $k_x$, $k_y$ and $k_{xy}$ are generalized curvatures. Equations (45) are expressed in matrix form by

$$\boldsymbol{\varepsilon}_n = \begin{bmatrix} \varepsilon_x & \varepsilon_y & \gamma_{xy} \end{bmatrix}^T = \mathbf{p}^T \mathbf{e} \qquad (46)$$

In (46), the position matrix $\mathbf{p}$ and the generalized strain vector $\mathbf{e}$ are defined by

$$\mathbf{p} = \begin{bmatrix} 1 & z & 0 & 0 & 0 & 0 \\ 0 & 0 & 1 & z & 0 & 0 \\ 0 & 0 & 0 & 0 & 1 & z \end{bmatrix}^T \qquad (47)$$

$$\mathbf{e} = \begin{bmatrix} e_x & k_x & e_y & k_y & e_{xy} & k_{xy} \end{bmatrix}^T \qquad (48)$$

The derivatives of the longitudinal strains $\varepsilon_x$, $\varepsilon_y$ and $\gamma_{xy}$ with respect to $x$ and $y$ are

determined by

$$\boldsymbol{\varepsilon}'_n = \mathbf{p}^T \mathbf{e}' \qquad \dot{\boldsymbol{\varepsilon}}_n = \mathbf{p}^T \dot{\mathbf{e}} \qquad (49)a,b$$

where the derivatives $\mathbf{e}'$ and $\dot{\mathbf{e}}$ of the generalized strain vector are expressed by

$$\mathbf{e}' = \begin{bmatrix} e'_x & k'_x & e'_y & k'_y & e'_{xy} & k'_{xy} \end{bmatrix}^T \qquad \dot{\mathbf{e}} = \begin{bmatrix} \dot{e}_x & \dot{k}_x & \dot{e}_y & \dot{k}_y & \dot{e}_{xy} & \dot{k}_{xy} \end{bmatrix}^T \qquad (50)a,b$$

Substituting (49) in (44) yields

$$s_{xz}(z) = \mathbf{S}_x^T(z)\,\mathbf{e}' + \mathbf{S}_{xy}^T(z)\,\dot{\mathbf{e}} \qquad s_{yz}(z) = \mathbf{S}_{xy}^T(z)\,\mathbf{e}' + \mathbf{S}_y^T(z)\,\dot{\mathbf{e}} \qquad (51)a,b$$

where $\mathbf{S}_x(z)$, $\mathbf{S}_y(z)$ and $\mathbf{S}_{xy}(z)$ are vectors defined by

$$\mathbf{S}_x^T(z) = -\int_{z_A}^{z} \mathbf{D}_x \mathbf{p}^T \mathrm{d}z \qquad \mathbf{S}_y^T(z) = -\int_{z_A}^{z} \mathbf{D}_y \mathbf{p}^T \mathrm{d}z \qquad \mathbf{S}_{xy}^T(z) = -\int_{z_A}^{z} \mathbf{D}_{xy} \mathbf{p}^T \mathrm{d}z \qquad (52)a\text{-}c$$

The vector $\mathbf{F}$ of generalized stresses per unit length is expressed by

$$\mathbf{F} = \begin{bmatrix} N_x & M_x & N_y & M_y & N_{xy} & M_{xy} \end{bmatrix}^T = \int_{z_A}^{z_B} \mathbf{p}\,\mathbf{s}_n\,\mathrm{d}z \qquad (53)$$

The derivatives $\mathbf{F}'$ and $\dot{\mathbf{F}}$, with respect to $x$ and $y$, are determined by

$$\mathbf{F}' = \begin{bmatrix} N'_x & M'_x & N'_y & M'_y & N'_{xy} & M'_{xy} \end{bmatrix}^T = \int_{z_A}^{z_B} \mathbf{p}\,\mathbf{s}'_n\,\mathrm{d}z$$

$$\dot{\mathbf{F}} = \begin{bmatrix} \dot{N}_x & \dot{M}_x & \dot{N}_y & \dot{M}_y & \dot{N}_{xy} & \dot{M}_{xy} \end{bmatrix}^T = \int_{z_A}^{z_B} \mathbf{p}\,\dot{\mathbf{s}}_n\,\mathrm{d}z \qquad (54)a,b$$

Substituting (41) and (49) in (54) yields

$$\mathbf{F}' = \mathbf{K}\,\mathbf{e}' \qquad \dot{\mathbf{F}} = \mathbf{K}\,\dot{\mathbf{e}} \qquad (55)a,b$$

where the stiffness matrix $\mathbf{K}$ is defined by

$$\mathbf{K} = \int_{z_A}^{z_B} \mathbf{p}\,\mathbf{D}\,\mathbf{p}^T\,\mathrm{d}z \tag{56}$$

Equations (51) and (55) yield the shear stresses $s_{xz}$ and $s_{yz}$. These simple final equations are conceptually identical to the equivalent section method for beams, proposed by Diaz (1980) and Diaz and Schulz (1981). Equations (55) define two systems of six generalized stresses by six generalized strains. The derivatives of generalized stresses in (54), which can be directly evaluated from analytical solutions and finite element analyses, must satisfy the differential equilibrium equations (38).

**HIGHER-ORDER APPROXIMATIONS**

Although the first-order approximation is considered adequate for practical applications, a higher-order solution is also discussed. Deriving (35), substituting (33), and integrating, yield

$$s'_{xz}(z) = -\dot{s}_{yz}(z) = -\frac{1}{2}\int_{z_A}^{z}\left(s''_x - \ddot{s}_y\right)\mathrm{d}z$$

$$\dot{s}_{xz}(z) = -\int_{z_A}^{z}\left(\dot{s}'_x + \ddot{s}_{xy}\right)\mathrm{d}z \tag{57}a\text{-}c$$

$$s'_{yz}(z) = -\int_{z_A}^{z}\left(s''_{xy} + \dot{s}'_y\right)\mathrm{d}z$$

The second-order approach is defined by neglecting the second derivatives of the shear stresses $s_{xz}$ and $s_{yz}$. Considering that the material properties do not significantly vary in $x$ and $y$, the following approximation is adopted:

$$\mathbf{s}''_n = \mathbf{D}\boldsymbol{\varepsilon}''_n \qquad \ddot{\mathbf{s}}_n = \mathbf{D}\ddot{\boldsymbol{\varepsilon}}_n \qquad \dot{\mathbf{s}}'_n = \mathbf{D}\dot{\boldsymbol{\varepsilon}}'_n \tag{58}a\text{-}c$$

Using partition (42) and substituting (58) in (57) yields

$$s'_{xz}(z) = -\dot{s}_{yz}(z) = -\frac{1}{2}\int_{z_A}^{z}\left(\mathbf{D}_x\,\varepsilon''_n - \mathbf{D}_y\,\ddot{\varepsilon}_n\right)dz$$

$$\dot{s}_{xz}(z) = -\int_{z_A}^{z}\left(\mathbf{D}_x\,\dot{\varepsilon}'_n + \mathbf{D}_{xy}\,\ddot{\varepsilon}_n\right)dz \qquad (59)a\text{-}c$$

$$s'_{yz}(z) = -\int_{z_A}^{z}\left(\mathbf{D}_{xy}\,\varepsilon''_n + \mathbf{D}_y\,\dot{\varepsilon}'_n\right)dz$$

where $\varepsilon''_n = \mathbf{p}^T\mathbf{e}''$, $\ddot{\varepsilon}_n = \mathbf{p}^T\ddot{\mathbf{e}}$ and $\dot{\varepsilon}'_n = \mathbf{p}^T\dot{\mathbf{e}}'$. Deriving (54) and substituting (58) yields the derivatives $\mathbf{F}''$, $\ddot{\mathbf{F}}$ and $\dot{\mathbf{F}}'$ of the vector of generalized stresses, which are expressed by

$$\mathbf{F}'' = \mathbf{K}\,\mathbf{e}'' \qquad \ddot{\mathbf{F}} = \mathbf{K}\,\ddot{\mathbf{e}} \qquad \dot{\mathbf{F}}' = \mathbf{K}\,\dot{\mathbf{e}}' \qquad (60)a\text{-}c$$

where $\mathbf{K}$ is the stiffness matrix defined in (56).

Equations (59) and (60) evaluate the shear stress derivatives $s'_{xz}$, $\dot{s}_{xz}$, $s'_{yz}$ and $\dot{s}_{yz}$, formerly neglected by the first-order approximation. The partial derivatives $s''_{xz}$, $\ddot{s}_{xz}$, $\dot{s}'_{xz}$, $s''_{yz}$, $\ddot{s}_{yz}$ and $\dot{s}'_{yz}$, neglected in the second order approximation, can be considered by repeating the proposed approach in a third-order approximation. Higher-order approximations can be successively established in search of more precise results, but require the evaluation of additional generalized stresses. The first-order approximation is considered acceptable for reinforced concrete design, and it is recommended in the interest of simplifying the formulation.

**MATERIAL BEHAVIOR**

The proposed mechanical model can be implemented according to different constitutive formulations. The present analysis is based on the bidimensional constitutive model A proposed by Vecchio and Collins (1993). It uses the uniaxial stress-strain curve proposed by Popovics (1973), modified for high strength concrete by Thorenfeldt, Tomaszewicz and Jensen (1987) and calibrated by Collins and Porasz (1989). In model A, the softening effect in tension-compression state is expressed as a function of the ratio $\varepsilon_1/\varepsilon_2$, where $\varepsilon_1$ and $\varepsilon_2$ are respectively the principal tensile and compressive strains. In the present paper, this estimate is lower bounded by the softening coefficient formerly proposed by the same authors (1986), which is a function of $\varepsilon_1$. The tension-stiffening effect is as represented by Polak and Vecchio (1993), where concrete average tensile stresses, transmitted across the cracks, are limited by the reserve capacity of the reinforcement. Tension stiffening effect is assumed to be limited to a volume of concrete within 7.5 bar diameters from the reinforcement center, and the maximum shear stress at a crack is verified as recommended by Vecchio and Collins (1986). The shear slip at crack surfaces, Poisson's ratio and other secondary effects are not considered.

The tridimensional constitutive model uses the same uniaxial curves, but the principal strains $\varepsilon_1$ and $\varepsilon_2$ are replaced, in the same softening functions, by the strain parameters $\varepsilon'_1$ and $\varepsilon'_2$. Kirschner and Collins (1986) define $\varepsilon'_1$ as the square root of the sum of the squares of all positive principal strains. The strain $\varepsilon'_2$ is assumed as the minimum principal strain. The terms $E_{ij}$ in (24) are determined by

$$E_{ij} = \frac{\partial \sigma_i}{\partial \varepsilon_j} + \frac{\partial \sigma_i}{\partial \beta}\left(\frac{\partial \beta}{\partial \varepsilon'_1}\frac{\partial \varepsilon'_1}{\partial \varepsilon_j} + \frac{\partial \beta}{\partial \varepsilon'_2}\frac{\partial \varepsilon'_2}{\partial \varepsilon_j}\right) \qquad (61)$$

where $\beta(\varepsilon'_1,\varepsilon'_2)$ is the tension-softening coefficient. The partial derivatives in (61) are analytically presented by Santisi d'Avila (2008). Since $E_{ij} \neq E_{ji}$ under tension-compression states, matrices **E**, **C**, **D** and **K** are not necessarily symmetric.

**IMPLEMENTATION PROCEDURES**

The following procedure yields strains and stresses for a given set of internal forces:

1. An iteration is started considering a generalized strain vector **e** (48) and distributions of shear stresses $s_{xz}$ and $s_{yz}$ along the thickness. The first approximations can be zero.

2. At each layer, the vectors of in-plane strains $\boldsymbol{\varepsilon}_n$ (46) and transverse stresses $\mathbf{s}_t$ are known. Vectors $\mathbf{s}_n$ and $\boldsymbol{\varepsilon}_t$ are determined using a secondary iterative process based on (39), the final step of which yields the reduced constitutive matrix **D** (42).

3. The vector of generalized stresses **F** (53) is integrated. The procedure stops when the residual $\Delta \mathbf{F}$, between applied and resisting internal forces **F**, is considered relatively small.

4. The stiffness matrix **K** (56) and the derivatives of generalized strains $\mathbf{e}'$ and $\dot{\mathbf{e}}$ (50), evaluated by equations (55), yield a new approximation of the shear stresses $s_{xz}$ and $s_{yz}$ (51). Solving $\Delta \mathbf{F} = \mathbf{K}\, \Delta \mathbf{e}$ yields the strain increment $\Delta \mathbf{e}$ and a new approximation of the generalized strain vector **e**, restarting the main iterative process.

The simultaneous update of generalized strains and shear stresses proves to be numerically efficient, although each update assumes that the other parameter is restrained. The numerical efficiency is not affected by solving non-symmetric systems with low order matrices. The secondary iterative process, based on (39), is a Newton-Raphson procedure that solves

$$\begin{bmatrix} \Delta \mathbf{s}_n \\ \Delta \boldsymbol{\varepsilon}_t \end{bmatrix} = \begin{bmatrix} \mathbf{C}_{nn} - \mathbf{C}_{nt}\, \mathbf{C}_{tt}^{-1}\, \mathbf{C}_{tn} & \mathbf{C}_{nt}\, \mathbf{C}_{tt}^{-1} \\ -\mathbf{C}_{tt}^{-1}\, \mathbf{C}_{tn} & \mathbf{C}_{tt}^{-1} \end{bmatrix} \begin{bmatrix} \Delta \boldsymbol{\varepsilon}_n \\ \Delta \mathbf{s}_t \end{bmatrix} \qquad (62)$$

The ultimate load capacity is evaluated by another incremental procedure based on the arc-length method. The shear stresses are determined using the stiffness matrix **K** of the previous incremental step. Additional details are presented by Santisi d'Avila (2008).

**COMPARISON WITH TEST DATA REPORTED IN THE LITERATURE**

The proposed theory and the material model used are verified with experimental data obtained from the literature.

Vecchio and Collins (1982) analyze the response of membrane elements submitted to combined in-plane shear and normal stresses. Specimens PV, with 890 x 890 x 70 mm, are typically reinforced with two layers of welded wire mesh, which are heat-treated to exhibit a long yielding plateau. The selection presented in Table 1 excludes panels with concrete voids, panels which fail prematurely because of pull-out of shear connecting keys, and panels that require additional information on steel response to be properly analyzed under strain-hardening. Table 1 presents the main material and geometrical information, including the steel yielding stresses ($f_{yl}$, $f_{yt}$) and reinforcement ratios ($\rho_l$, $\rho_t$), which are defined, respectively, in each orthogonal direction. The cracking tensile strength of concrete is estimated by $0.33\sqrt{f_c}\,(\text{MPa})$. Most specimens are subjected to increasing pure shear, and $\tau_u$ designates the ultimate shear stress. Panels PV23, PV25 and PV28 combine shear stresses, denoted by $\tau$, with biaxial normal stresses, which are respectively equal to $-0.39\,\tau$, $-0.69\,\tau$ and $+0.32\,\tau$. The biaxial compressive stresses of specimen PV29 are expressed in modulus by $\tau - 3.80\,\text{MPa}$, when $\tau > 3.80\,\text{MPa}$.

The ultimate loads predicted by Vecchio and Collins (1982), Bentz, Vecchio and Collins (2006) and the proposed procedure show good correlation. The contribution of tension-stiffening is analyzed by excluding and including this effect. The predicted results are denoted respectively by PT (pure tension) and TS (tension stiffening), and ultimate loads are not significantly different.

Polak and Vecchio (1993) investigate the behavior of shell elements subjected to biaxial bending and in-plane forces, using a shell element tester that is capable of applying uniform load conditions. Specimen SM4 is of special interest because it presents symmetrical top and bottom reinforcement in two orthogonal directions, orientated at $45°$ with respect to the applied loads $M$ and $P$ (Figure 4). The reinforcement ratios are 1.32% ($\phi 20$ spaced at 0.076m) and 0.44% ($\phi 10$ spaced at 0.076m), per layer, in each direction. The corresponding stress-strain curves present definite plateaus at 425 and 430 MPa, strain hardening after 12 and 20 mm/m and ultimate strength of 611 and 480 MPa. The effective test dimensions are 1524 x 1524 x 316 mm, the concrete compression peak stress is 64 MPa at 2.6 mm/m, and the split cylinder tensile test yields 2.76 MPa. The final load level is reported as $M_u = 205 \text{ kNm/m}$ and $P_u = 820 \text{ kN/m}$.

The element $x$-axis is defined in the direction of the stronger reinforcement. The experimental surface strains $\varepsilon_y$ and reinforcement strains $\varepsilon_{sx}$ and $\varepsilon_{sy}$ are compared to the theoretical results (Figure 4). The predicted strains, excluding the concrete tensile strength (NT - no tension) and including the tension stiffening (TS), describe tension effects at intermediate load levels and demonstrate good correlation.

The predicted and observed ultimate moments $M_u$ are discussed in Table 2 according to three conditions: excluding the concrete tensile strength (NT), taking into account the concrete tensile strength but excluding the tension-stiffening effect (PT) and including both concrete tensile strength and tension-stiffening effect (TS). All assumptions predict conservative results in agreement with the observed data. In this analysis, ultimate loads are not influenced by tensile effects.

The predicted stresses and strains, neglecting concrete tensile strength, are presented at a load level near to failure (Figure 5). The stress on the weaker reinforcement increases because of strain hardening, but it is still inferior to the ultimate strength. Concrete reaches its reduced compressive strength ($\sigma = -\beta f_c$), as shown in Figure 6.

Adebar (1989) and Adebar and Collins (1994), using the same element tester, analyze 1524 x 1524 mm shell elements subjected to in-plane and transverse shear loads. The specimens selected for this verification have very small amounts of shear reinforcement (0.08% reinforcement ratio). Some other specimens, which present larger amounts of shear reinforcement, are discarded because they do not fail during the test. The specimens denoted by SP3, SP4, SP7, SP8 and SP9 are 310 mm thick and have large amounts of symmetrical in-plane reinforcement, equally arranged in orthogonal grids oriented at 45° with respect to their sides. The stress-strain curves show definite plateaus and the strain hardening response is described by Adebar (1989). The yield and ultimate stresses of the shear reinforcement are respectively 460 MPa and 570 MPa and other details are presented in Table 3. In the present numerical analyses, the cracking strength of concrete is estimated based on the results of the split test. The specimens are subjected to different load combinations, which are proportionally increased during the testing procedure. The in-plane and transverse forces are constant along the specimen, but bending moments vary linearly (Figure 7). The $xy$ coordinate system is associated with the in-plane reinforcement. Generic transverse shear forces $V_{x'} = \sqrt{2}$ kN/m and $V_{y'} = 0$ kN/m, applied in the $x'y'$ coordinate system, correspond to $V_x = V_y = 1$ kN/m. The derivatives $M'_x$, $M'_y$, $M'_{xy}$, $\dot{M}_x$, $\dot{M}_y$ and $\dot{M}_{xy}$ are equal to $1/2$ kN/m. The load capacity is verified at $x' = \pm 0.75$ m, transforming the corresponding bending moment $M_{x'}$ to $xy$ coordinates.

Theoretical and observed stirrup strains are compared in Figure 8. The reference shear stress $V/t$ is defined by $V/t = V_x/t = V_y/t$, where $t$ is the thickness of the shell. The experimental transverse strains cannot be exactly represented by the proposed formulation, since they are measured with strain gauges on unbounded stirrups. According to hypothesis 2, no slip is considered between concrete and reinforcing bars. In spite of this limitation, predicted average stirrup strains are compared, in magnitude, excluding and including the tension stiffening effect. The theoretical strains of specimen SP4, including the tension stiffening effect, show low levels of stress. The same results are predicted by Adebar and Collins (1994). The necessary stirrup contribution to the equilibrium, between cracks, justifies the large experimental strains of the unbounded stirrups.

Although the properties of specimens SP3, SP4, SP7, SP8 and SP9 are not exactly the same, it is possible to establish a relationship between transverse shear capacity and membrane shear forces (Figure 9). The ultimate loads predicted by the proposed theory are compared to other formulations (Polak and Vecchio 1993; Adebar and Collins 1994), and give a better approximation. Results excluding tension-stiffening effects are significantly lower. The reduction of longitudinal strains increases the shear capacity considerably.

The response of specimen SP7 is evaluated close to failure, including the concrete tensile strength and the tension-stiffening effect (Figure 10). Minute negative shear stresses $\tau_{yz}$ are detected in the tensile cover region. They are associated with the negative derivatives of the stress-strain diagram for concrete under tension, the descending branch of which represents declining tension stiffening as cracking progresses. The stirrups reach yield stress through an extensive area of the thickness. The principal compressive stresses $\sigma_2$ and $\sigma_3$ are compared with the concrete capacity in Figure 11, where the reduced concrete strength due to compression softening is defined by $\beta f_c$.

The spatial orientation of the principal compressive stresses close to failure, as predicted by the proposed theory, is presented in Figure 12. The orientation of the struts is more detailed for specimen SP7 (Figure 13). The direction of the shear struts, in the central core of the shell, follows the direction of the principal transverse shear force.

**CONCLUSIONS**

The proposed mechanical model yields through-the-thickness distributions of stresses and strains in reinforced concrete shells, considering the simultaneous effect of in-plane and transverse loads. The comparison with test data reported in the literature confirms that the formulation is able to accurately model in-plane conditions, flexural behavior and transverse shear effects. The constitutive model adopted predicts results which are in agreement with the observed data.

The proposed theory provides a promising solution for the design of reinforced concrete shells. A practical design method can be developed by applying optimization techniques to balance the internal forces with a minimum amount of total reinforcement. Strain limitations can be considered as complementary optimization conditions. The bipartition method, starting from the minimum and maximum code requirements, is a simple approach that is sufficient for fixed reinforcement ratios.

However, the influence of tension effects demands additional investigation. Concrete tension effects are often neglected in reinforced concrete design, because they are usually considered to have little to no influence on the ultimate load capacity. This hypothesis is confirmed in PV and SM specimens (membrane and shell elements subjected to in-plane conditions). The fact that the load capacity of SP specimens is significantly affected by tension stiffening shows that this approach may be too conservative for shells with transverse shear forces, particularly when dealing with large amounts of flexural reinforcement, small amounts of shear reinforcement and monotonic loadings.

The verification of tension effects, with additional experimental data, and the implementation of finite element formulations, based on the proposed mechanical model, call for further research.

**ACKNOWLEGMENTS**

The authors thank Prof. Benjamin Ernani Diaz and Prof. Giuseppe Ricciardi for their support. The second author acknowledges the doctoral fellowship granted by the Department of Civil Engineering of Messina University.

**APPENDIX I. REFERENCES**

Adebar, P., and Collins, M. P. (1994). "Shear design of concrete offshore structures." *ACI Struct. J.*, 91(3), 324-335.

Adebar, P. E. (1989). "Shear design of concrete offshore structures," PhD thesis, Univ. of Toronto, Toronto, Canada.

Armer, G. S. T. (1968). "Discussion of the reinforcement of slabs in accordance with a predetermined field of moments." *Concrete*, 2(8), 319-320.

Baumann, T. (1972a). "Tragwirkung orthogonaler Bewehrungsnetze beliebiger Richtung in Flächentragwerken aus Stahlbeton." *Deutscher Ausschuss für Stahlbeton*, 217.

Baumann, T. (1972b). "Zur Frage der Netzbewehrung von Flächentragwerken." *Bauingenieur*, 47(10), 367-377.

Bentz, E. C., Vecchio, F. J., and Collins, M. P. (2006). "Simplified modified compression field theory for calculating shear strength for reinforced concrete elements." *ACI Struct. J.*, 103(4), 614-624.

Collins, M. P., and Porasz, A. (1989). "Shear design for high strength concrete." *Bul. d'Information No. 193: Design aspects of high strength concrete*, CEB, Lausanne, Switzerland, 77-83.

Comité Euro-International du Béton, and Fédération Internationale de la Précontrainte. (1982). *Bul. d'Information No. 141: CEB/FIP manual on bending and compression*, Construction Press, London, U.K.

Cook, R. D., Malkus, D. S., and Plesha, M. E. (1989). *Concepts and applications of finite element analysis*, 3rd Ed., J. Wiley & Sons, New York, N.Y.

Diaz, B. E. (1980). "Dimensionamento à esforço cortante." *Estrutura*, 92, 36-54.

Diaz, B. E., and Schulz, M. (1981). "Design of reinforced concrete based on mechanics." *Reports of the Working Comissions No. 34: IABSE Col. Delft 1981*, IABSE, Zürich, Switzerland, 531-

544.

Falconer, B. H. (1956). "Theory of the stresses induced in reinforced concrete by applied two-dimensional stress." *ACI J.*, 53(9), 277-293.

Kirschner, U., and Collins, M. P. (1986). "Investigating the behavior of reinforced concrete shell elements." *Publ. No. 86-9*, Dept. of Civ. Engrg., Univ. of Toronto, Toronto, Canada.

Marti, P. (1990). "Design of concrete slabs for transverse shear." *ACI Struct. J.*, 87(2), 180-190.

Mitchell, D., and Collins, M. P. (1974). "Diagonal compression field theory - a rational model for structural concrete in pure torsion." *ACI J.*, 71(8), 396-408.

Nielsen, M. P. (1964). "Yield conditions for reinforced concrete shells in the membrane state." *Non Classical Shell Problems: IASS Symp. (1963: Warsaw, Poland)*, W. Olszak, ed., North-Holland, Amsterdam, The Netherlands, 1030-1040.

Polak, M. A., and Vecchio, F. J. (1993). "Nonlinear analysis of reinforced concrete shells." *Publ. No. 93-03*, Dept. of Civ. Engrg., Univ. of Toronto, Toronto, Canada.

Popovics, S. (1973). "A numerical approach to the complete stress-strain curve of concrete." *Cem. Concr. Res.*, 3(5), 583-599.

Santisi d'Avila, M. P. (2008). "Design of reinforced concrete structures using the equivalent section method," PhD thesis, Univ. of Messina, Messina, Italy.

Schnobrich, W. C. (1977). "Behavior of reinforced concrete structures predicted by the Finite Element Method." *Computer and Structures*, 7, 365-376.

Schulz, M. (1984). "Design of reinforced concrete plates and shells." *Proc., Int. Conf. on Structural Analysis and Design of Nuclear Power Plants*, Porto Alegre, Brazil.

Schulz, M. (1988). "Verificação geral de peças de concreto armado baseada no modelo da chapa fissurada," PhD thesis, Federal University of Rio de Janeiro, Rio de Janeiro, Brazil.

Stevens, N. J., Uzumeri, S. M., and Collins, M. P. (1987). "Analytical modelling of reinforced concrete subjected to monotonic and reversed loadings." *Publ. No. 87-01*, Dept. of Civ. Engrg., Univ. of Toronto, Toronto, Canada.

Thorenfeldt, E., Tomaszewicz, A., and Jensen, J. J. (1987). "Mechanical properties of high-strength concrete and application in design." *Proc., Int. Symp. on Utilization of High-Strength Concrete*, Tapir, Stavanger, Norway, 149-159.

Vecchio, F. J., and Collins, M. P. (1982). "The response of reinforced concrete to in-plane shear and normal stresses." *Publ. No. 82-03*, Dept. of Civ. Engrg., Univ. of Toronto, Toronto, Canada.

Vecchio, F. J., and Collins, M. P. (1986). "The modified compression field theory for reinforced concrete elements subjected to shear." *ACI J.*, 83(2), 219-231.

Vecchio, F. J., and Collins, M. P. (1993). "Compression response of cracked reinforced concrete." *J. Struct. Engrg.*, 119(2), 3590-3610.

Willam, K., Pramono, E., and Sture, S. (1987). "Fundamental issues of smeared crack models." *Fracture of concrete and rock: SEM-RILEM Int. Conf. (1987: Houston, Texas)*, S. P. Shah and S. E. Swartz, eds., Springer-Verlag, New York, N.Y., 142-157.

Wood, R. H. (1968). "The reinforcement of slabs in accordance with a pre-determined field of moments." *Concrete*, 2(2), 69-76.

Zhu, R. R. H., Hsu, T. T. C., and Lee, J. Y. (2001). "Rational shear modulus for smeared-crack analysis of reinforced concrete." *ACI Struct. J.*, 98(4), 443-450.

**APPENDIX II. NOTATION**

The following symbols are used in this paper:

$( )'$ = derivative of function with respect to variable $x$ $(\partial/\partial x)$

$(\dot{\ })$ = derivative of function with respect to variable $y$ $(\partial/\partial y)$

$( )^*$ = derivative of function with respect to variable $z$ $(\partial/\partial z)$

$\mathbf{C}$ = tangent constitutive matrix of reinforced concrete $(6\times 6)$

$\mathbf{D}$ = reduced constitutive matrix of reinforced concrete $(3\times 3)$

$\mathcal{E}$ = strain tensor

$\mathbf{E}$ = tangent constitutive matrix of concrete

$\mathbf{E}_s$ = tangent constitutive matrix of reinforcement

$E_{ij}$ = elasticity moduli of concrete $(i, j = 1, 2, 3)$

$E_{sx}, E_{sy}, E_{sz}$ = elasticity moduli of reinforcement

$\mathbf{e}$ = vector of generalized strains

$e_x, e_y, e_{xy}$ = axial strains and in-plane shear strain at $z = 0$

$\mathbf{F}$ = vector of generalized stresses per unit length

$f_c$ = concrete strength in compression

$f_y$ = steel yielding stress

$f_u$ = steel ultimate stress

$G_{ij}$ = tangent shear moduli $(i, j = 1, 2, 3; \text{with } i \neq j)$

$\mathbf{I}$ = identity matrix

$\mathbf{K}$ = stiffness matrix of the reinforced concrete shell element

$k_x, k_y, k_{xy}$ = curvatures of the cross-section

$N_x, N_y, N_{xy}$ = axial forces and in-plane shear force per unit length

$M_u$ = ultimate bending moment

$M_x, M_y, M_{xy}$ = bending and twisting moments per unit length

$\mathbf{p}$ = position matrix

$P_u$ = ultimate axial force

$S$ = concrete stress tensor

$\mathbf{S}_x, \mathbf{S}_y, \mathbf{S}_{xy}$ = vectors of stiffness properties of reinforced concrete between $z_A$ and $z$

$\mathbf{s}$ = vector of analogous stresses in reinforced concrete

$s_x, s_y, s_{xy}$ = in-plane analogous stresses in reinforced concrete

| | | |
|---|---|---|
| $s_{yz}, s_{xz}, s_z$ | = | transverse analogous stresses in reinforced concrete |
| $\mathbf{T}$ | = | rotation matrix |
| $t$ | = | shell thickness |
| $V_u$ | = | ultimate shear force |
| $V_x, V_y$ | = | transverse shear forces per unit length |
| $x, y, z$ | = | coordinate axes |
| $x_1, x_2, x_3$ | = | principal coordinate axes |
| $z_A, z_B$ | = | $z$-coordinates of the cross-section upper and lower bounds |
| $\beta$ | = | softening coefficient |
| $\gamma_{xy}, \gamma_{xz}, \gamma_{yz}$ | = | shear strains |
| $\gamma_{12}, \gamma_{23}, \gamma_{13}$ | = | shear strains in principal coordinates |
| $\boldsymbol{\varepsilon}$ | = | strain vector |
| $\varepsilon_{sx}, \varepsilon_{sy}, \varepsilon_{sz}$ | = | reinforcement strains |
| $\varepsilon_{sx0}, \varepsilon_{sy0}, \varepsilon_{sz0}$ | = | residual strain of prestressed tendons |
| $\varepsilon_x, \varepsilon_y, \varepsilon_z$ | = | axial strains |
| $\varepsilon_{xy}, \varepsilon_{xz}, \varepsilon_{yz}$ | = | shear terms of the strain tensor |
| $\varepsilon_1, \varepsilon_2, \varepsilon_3$ | = | maximum, intermediate and minimum principal strains |
| $\varepsilon_1', \varepsilon_2'$ | = | strain parameters |
| $\boldsymbol{\rho}_s$ | = | matrix of reinforcement ratios |
| $\rho_{sx}, \rho_{sy}, \rho_{sz}$ | = | reinforcement ratios |
| $\boldsymbol{\sigma}$ | = | concrete stress vector |
| $\boldsymbol{\sigma}_s$ | = | reinforcement stress vector |
| $\sigma_{sx}, \sigma_{sy}, \sigma_{sz}$ | = | reinforcement stresses |
| $\sigma_x, \sigma_y, \sigma_z$ | = | concrete axial stresses |
| $\sigma_1, \sigma_2, \sigma_3$ | = | concrete axial stresses in principal coordinates |
| $\tau_{xy}, \tau_{yz}, \tau_{xz}$ | = | concrete shear stresses |
| $\tau_{12}, \tau_{23}, \tau_{13}$ | = | concrete shear stresses in principal coordinates |
| $\boldsymbol{\Phi}$ | = | rotation matrix |
| $\phi_{ij}$ | = | terms of matrix $\boldsymbol{\Phi}$ $(i = 1, 2, 3; \ j = x, y, z)$ |